# An improved method to access initial states in relativistic heavy-ion collisions

Somadutta Bhatta[1], Vipul Bairathi[2,a]

[1] Department of Chemistry, Stony Brook University, Stony Brook, NY 11794, USA
[2] Instituto de Alta Investigación, Universidad de Tarapacá, Casilla 7D, Arica 1000000, Chile

**Abstract** Observables in heavy-ion collisions are generally categorized into centralities, which reflect an average over events within a range of impact parameters including a wide variety of initial-state configurations. A multiple binning method using spectator neutrons within each centrality has been shown to provide access to events with rare initial-state conditions. This work suggests an improvement in quantifying the difference between standard centrality and spectator neutron binning towards accessing the initial-state properties. A selection of events with higher initial-state density at a fixed participating nucleon number was observed to result in larger final-state particle production and smaller elliptic flow. The relative difference between observables in centrality and spectator binning shows reduced sensitivity for the observables dominated by impact parameter fluctuations in the initial state, such as triangular flow. This property renders the spectator binning method a good candidate for separating geometric contributions from random fluctuations in the initial state towards final-state observables.

## 1 Introduction

In relativistic heavy-ion collisions, a large amount of energy deposited in the collision region deconfines the hadronic matter into a strongly interacting medium called quark–gluon plasma (QGP) [1,2]. Subsequent cooling and expansion of the QGP leads to the production of particles, which carry important information about the initial state of the collision. Initial states of heavy-ion collisions are very important and pose one of the greatest challenges in understanding the properties and dynamics of the QGP medium produced in relativistic heavy-ion collisions. State-of-the-art Bayesian analyses are generally used to constrain the transport properties of the QGP. However, large uncertainty in these analyses arises from the specific assumptions on the initial state prior to hydrodynamic expansion [3–5]. For example, depending on the choice of initial conditions in modeling of heavy-ion collisions, the extracted values of the shear viscosity to entropy density ratio at the Relativistic Heavy Ion Collider (RHIC) can vary by a factor of 2 [6–11].

One of the primary variables controlling the amount of energy deposited in the collision region is the impact parameter ($b$), which is the transverse distance between the center of two colliding nuclei. Collision events are categorized into "centralities" which correlate directly with $b$. In model calculations, the total number of participating nucleons is used to define the centrality, whereas the number of charged particles ($N_{ch}$) in a given pseudorapidity ($\eta$) region is used in experiments to define event centrality. In a simple Monte Carlo Glauber model simulation of heavy-ion collisions, initial-state observables (e.g., transverse overlap area ($S$), eccentricity of overlap region ($\varepsilon_n$), number of participating nucleons ($N_{part}$), and number of binary collisions ($N_{coll}$)) show monotonicity as a function of event centrality [12,13]. The pressure gradient in the initial state converts the spatial eccentricity to final-state momentum anisotropy. This final-state asymmetry is quantified by a Fourier expansion of azimuthal distribution of particles. The second and third coefficients in the Fourier expansion are called the elliptic ($v_2$) and triangular flow ($v_3$), respectively [14,15]. Existing transport models and hydrodynamic simulations describe the evolution of final-state observables as a function of centrality [16–20]. However, the current model predictions are dependent on the choice of initial state in heavy-ion collisions.

Estimates of initial-state parameters and measurements of final-state observables as a function of centrality only reflect the mean values over events within a centrality range. However, such an averaging procedure limits detailed access to the initial-state configurations contributing to the mean. The spectator nucleons ($N_{\text{spec}}$, defined as $N_{\text{part}}$ subtracted from the total number of nucleons) is an event property that also correlates strongly with the impact parameter and can be used to categorize events on a similar footing as $N_{\text{part}}$ or $N_{\text{ch}}$. Simply put, $N_{\text{spec}}$ in an event is inversely correlated with the $N_{\text{part}}$. Reference [21] showed that $N_{\text{spec}}$ carries the initial-state information, which can be used to select specific collision configurations such as body-tip in collisions of deformed uranium nuclei. A very similar "multi-differential" approach was recently used by the ALICE collaboration at CERN to disentangle the effects of effective energy and multiplicity towards production of strange hadrons [22].

[a] e-mail: vipul.bairathi@gmail.com (corresponding author)

Despite the strong correlation between $N_{\text{spec}}$ and $N_{\text{part}}$, the $N_{\text{ch}}$ has been shown to have a reduced correlation with the $N_{\text{spec}}$ stemming from interactions and scatterings in the overlap region [23]. The exact magnitude of the correlation between $N_{\text{spec}}$ and $N_{\text{ch}}$ might vary depending on the $\eta$ in which $N_{\text{ch}}$ is defined. The lower correlation between $N_{\text{ch}}$ and $N_{\text{spec}}$ forms a basis to use $N_{\text{spec}}$ as an approximately independent and new classifier of events. The advantage of such binning procedure in accessing the wide variety of initial states for events within any given centrality has already been studied and reported in Refs. [21,24]. In heavy-ion experiments, it is only possible to obtain information on the number of spectator neutrons. Therefore, we define $n_{\text{spec}}$ as the sum of neutron spectators moving in the left and right directions along the collision axis. In the Relativistic Heavy Ion Collider (RHIC) at Brookhaven National Laboratory (BNL), two zero-degree calorimeters (ZDCs) placed at $|\eta| > 6.0$ capture the spectator neutrons coming from the interaction region along the beam pipe [25,26]. Reference [25] reported the single neutron peak resolution of $\sigma_E/E \sim 40\%$ in peripheral collisions of Au+Au nuclei with current STAR ZDC, owing to clustering of a relatively larger number of neutrons. Therefore, we limit our measurements to a 0–60% centrality range in this study.

Before using this novel method to understand initial-state properties of heavy-ion collisions, it is important to develop a simple method that can differentiate between the extent of information on the initial state obtained from centrality classification and $n_{\text{spec}}$ classification of events. In addition, model simulations provide a better understanding of the behavior of observables when plotted as a function of the aforementioned event classifiers. Therefore, in this work, we used a transport model to compare observables between (i) a centrality event classifier based on $N_{\text{ch}}$ cuts ("centrality binning") and (ii) a multiple binning procedure in which events within a given centrality are further classified into several subgroups using $n_{\text{spec}}$ as an event classifier ("L+R binning").

This paper is structured as follows. Section 2 describes the justification for the new methodology used and suggested improvements in event classification. We present our results and discuss them in Sect. 3. Lastly, in Sect. 4, we summarize our findings and provide an outlook.

## 2 Methodology

A multi-phase transport (AMPT) [27] model in the string melting version is used to simulate collision events for this study. The AMPT model uses the same initial conditions as the HIJING model [28]. Zhang's parton cascade [29] follows to take into account partonic interactions which finally recombine with their parent strings that fragment into hadrons within the Lund string fragmentation model [30]. There is a final-stage hadronic afterburner before the hadron freeze-out. The AMPT model has been shown to reproduce experimentally measured observables such as collective flow and $dN_{\text{ch}}/d\eta$ to a good approximation [19,20]. In this work, $\sim$2 million Au+Au collision events at $\sqrt{s_{\text{NN}}} = 200$ GeV are simulated.

The multiple binning method used in this study follows Ref. [24]. We provide a brief outline of this procedure here. The centrality of an event is defined based on $N_{\text{ch}}$, which is calculated within $|\eta| < 0.5$. Based on the percentile of $N_{\text{ch}}$ distribution, events are classified into eight centrality bins each of 10 percentile as shown in Fig. 1a for Au+Au collisions at $\sqrt{s_{\text{NN}}} = 200$ GeV. The $n_{\text{spec}}$ distribution in each of the eight centralities is further divided into regular intervals of $n_{\text{spec}}$ as shown in Fig. 1b for 20–30% centrality. The $n_{\text{spec}}$ distribution displays a prominent peak around 125–140 and falls off rapidly on either side. The averaging over events within a centrality range in centrality binning reflects the properties of the events around the peak of total spectator neutron distribution. We can investigate the properties of the rare events with fewer or higher values of $n_{\text{spec}}$ compared with the central value with the introduction of the L+R binning.

Pearson correlators are widely used to quantify the correlations between any two observables in experimental heavy-ion collisions [23,31,32]. To emphasize the reduction in correlations between $N_{\text{ch}}$ and $n_{\text{spec}}$ in comparison with $N_{\text{part}}$ and $n_{\text{spec}}$, two Pearson correlators are defined on an event-by-event basis, $\rho(n_{\text{spec}}, N_{\text{part}})$ and $\rho(n_{\text{spec}}, N_{\text{ch}})$. Figure 1c shows a comparison between $\rho(N_{\text{spec}}, N_{\text{part}})$, $\rho(n_{\text{spec}}, N_{\text{part}})$, and $\rho(n_{\text{spec}}, N_{\text{ch}})$ as a function of centrality. The $N_{\text{spec}}$ is mathematically inversely correlated with $N_{\text{part}}$, which results in a $\rho(N_{\text{spec}}, N_{\text{part}})$ of $-1$, irrespective of centrality. However, because we use the number of spectator neutrons for the correlation measure (not the total number of spectators), and due to event-by-event fluctuations, the magnitude of $\rho(n_{\text{spec}}, N_{\text{part}})$ shows small deviations from the expected value of $-1$. In addition, as we go from central to peripheral centralities, the absolute magnitude of $\rho(n_{\text{spec}}, N_{\text{part}})$ is observed to decrease by about 10%,

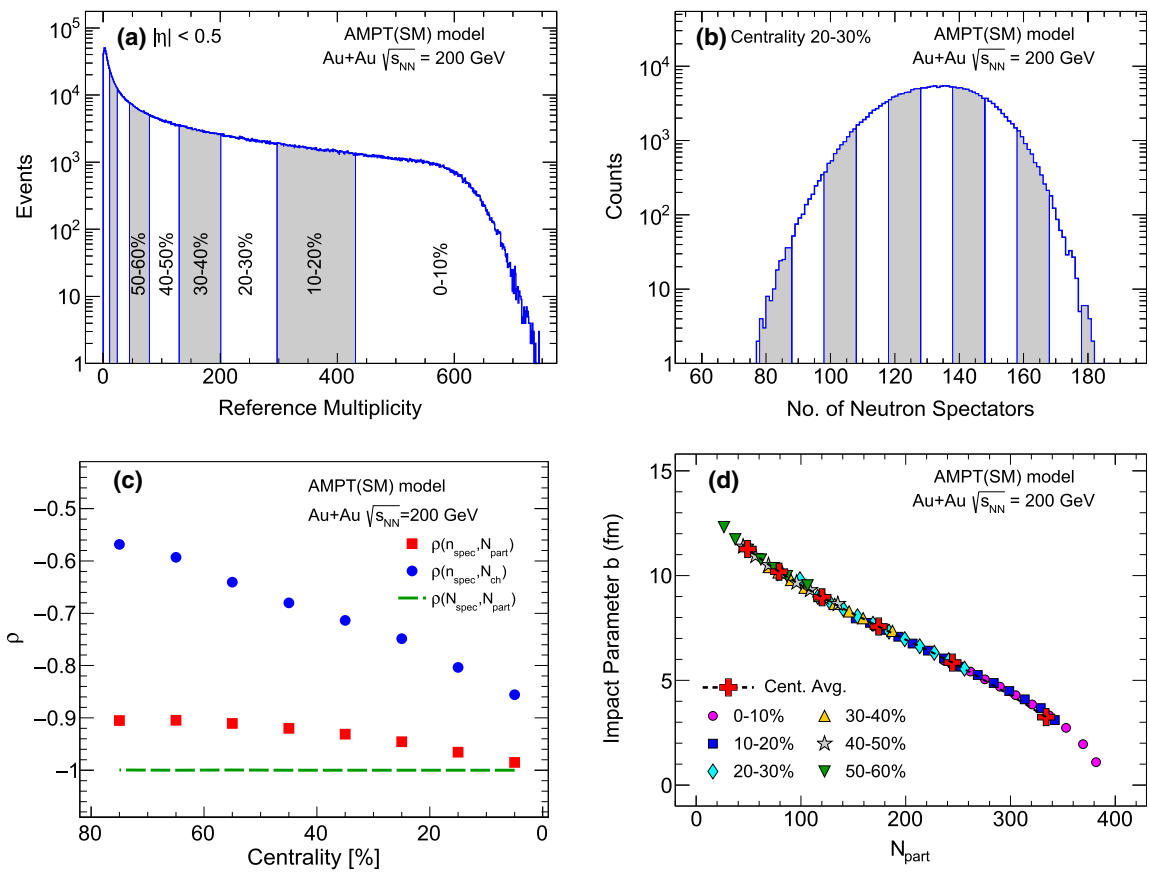

**Fig. 1** **a** The multiplicity in $|\eta| < 0.5$ ($N_{ch}$) for minimum-bias Au+Au collisions at $\sqrt{s_{NN}} = 200$ GeV. In alternating white and gray bands, different centrality bins from 0–10% to 70–80% are shown. **b** The total spectator neutron number ($n_{spec}$) distribution for 20–30% centrality. The different L+R bins are shown in alternating white and gray bands. **c** Pearson correlator between $N_{spec}$–$N_{part}$ (solid line), $n_{spec}$–$N_{part}$ (solid box), and $n_{spec}$–$N_{ch}$ (solid circle) as a function of centrality for Au+Au collisions at $\sqrt{s_{NN}} = 200$ GeV. **d** The impact parameter ($b$) vs. $N_{part}$ for different centrality and L+R bins

which might be due to increased fluctuation contributions to the spectator numbers with decreasing overlap area. In contrast, the absolute magnitude of $\rho(n_{spec}, N_{ch})$ is observed to decrease by about 45% going from central to peripheral events.

Because $N_{ch}$ and $n_{spec}$ are very weakly correlated, classifying events with $n_{spec}$ in addition to $N_{ch}$ is expected to give us increasingly more detail on event properties in relatively peripheral events. The Pearson correlator is equivalent to the "centered cosine similarity" between two random variables. Therefore, in $n_{spec}$–$N_{ch}$ parameter space, the cosine of the angle between centered $n_{spec}$ ($n_{spec}-\langle n_{spec} \rangle$) and centered $N_{ch}$ ($N_{ch}-\langle N_{ch} \rangle$) is given by $\rho(n_{spec}, N_{ch})$. Therefore, a decreasing magnitude of $\rho(n_{spec}, N_{ch})$ denotes a larger opening angle between the $n_{spec}$–$N_{ch}$ axes in $n_{spec}$–$N_{ch}$ parameter space, as shown in Fig. 1c. As a result, we obtain more complementary information about an event in the $n_{spec}$–$N_{ch}$ parameter space as we go from central to peripheral collision events using a multiple binning procedure. In real experiments, the exact magnitude of correlation between $N_{spec}$ and $dN_{ch}/d\eta$ might be different from that observed here, but the overall behavior is expected to remain the same. It might also be of interest to compare $\rho(n_{spec}, N_{ch})$ as a function of centrality between systems of different sizes to further understand its origin.

For centrality binning, observables are averaged over centrality intervals of 10% each. For L+R binning, observables are calculated by taking an average over equally spaced L+R bins within any given centrality. Figure 1d shows the variation in $b$ with $N_{part}$, in which centrality and L+R bins are observed to follow the same trend. This implies that the L+R bins in any given centrality can be used to select events with similar impact parameter and $N_{part}$ as the usual centrality binning. Thus, the introduction of $n_{spec}$ as an additional event classifier does not intrinsically select events with different centrality and thus does not bias our event selection. Therefore, any deviation from the centrality averaged trend, if observed in

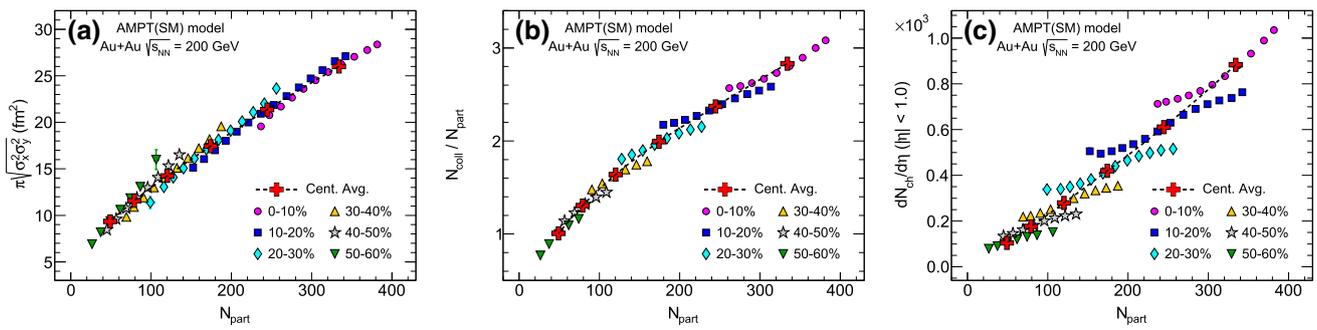

**Fig. 2** Plots for minimum-bias Au+Au collisions at $\sqrt{s_{NN}} = 200$ GeV. (Left) Overlap area vs. $N_{part}$; (middle) $N_{coll}/N_{part}$ vs. $N_{part}$; (right) $dN_{ch}/d\eta$ vs. $N_{part}$. Centrality average values are shown by the "+" symbol in red, whereas the values for L+R binning in each centrality are shown in solid circles of different colors

L+R binning, allows access to novel initial state configurations in heavy-ion collisions.

## 3 Results

This work focuses on the deeper insights gained from L+R binning in addition to centrality binning. The observables considered are $N_{part}$, $N_{coll}$, $N_{ch}$, $S$, $\varepsilon_n$, and $v_n$ for $n = 2, 3$. The transverse overlap area is calculated as $S = \pi\sqrt{\sigma_x^2 \sigma_y^2}$, where $\sigma_x$ and $\sigma_y$ are defined as $\sqrt{\langle x^2 \rangle - \langle x \rangle^2}$ and $\sqrt{\langle y^2 \rangle - \langle y \rangle^2}$, respectively, and $(x, y)$ denote the position of participating nucleons in the transverse plane. Eccentricities $\varepsilon_n = -\langle r^n e^{-in\phi} \rangle / \langle r^n \rangle$ are calculated from the positions $(r, \phi)$ of participating nucleons in the transverse plane. The coordinate system is shifted to the center of mass of participating nucleons. The calculation of the initial-state observables closely follows Ref. [33]. To estimate the final-state flow coefficients, the event-plane method is used [14]. In this method, the reaction-plane angle is estimated by the event-plane angle for each harmonic, and the azimuthal anisotropy of particles is calculated about this plane [14].

Figure 2a shows $S$ as a function of $N_{part}$. For centrality averaged values, the initial overlap area increases from peripheral to central collisions due to the decrease in the impact parameter. The correlation between $S$ and $N_{part}$ is different between L+R binning and centrality binning. In L+R binning, $S$ rises faster than the centrality averaged trend. The difference between trends shown by L+R and centrality binning is dependent on event centrality. Using L+R bins, it is possible to access events with a wider range of $S$ and $N_{part}$ compared with centrality averaged values in any given centrality. In addition, at a given $N_{part}$, we could choose events with different $S$ using L+R binning, which could help in studying the impact of initial participating nucleon density on final-state observables. At a fixed $N_{part}$, the area in L+R binning for more central events is observed to be smaller than that of a comparatively peripheral centrality. Thus, multiple binning procedure allows us to establish that for a given $N_{part}$, events in central collisions are more densely packed than relatively peripheral collisions. Therefore, the properties of events with different initial $N_{part}$ densities could be studied using the multiple binning procedure.

Figure 2b shows $N_{coll}/N_{part}$ as a function of $N_{part}$. $N_{coll}/N_{part}$ represents the number of binary collisions per participating nucleon. For a denser overlap area, a single $N_{part}$ is expected to encounter a larger number of binary collisions, and hence should have larger $N_{coll}$. This is clearly observed in Fig. 2b. For a given $N_{part}$, more central events in L+R binning have a larger number of binary collision per $N_{part}$. Such rare events with the same $N_{part}$ but very different interaction rates inside the collision region could help disentangle the effect of the initial state on measured final-state observables. Such a selection of events could not otherwise be accessed using the standard centrality binning alone.

Figure 2c shows $dN_{ch}/d\eta$ as a function of $N_{part}$. The $dN_{ch}/d\eta$ is expected to increase with $N_{part}$. A larger number of binary collisions per participant nucleon would lead to a larger value of $dN_{ch}/d\eta$ according to a simple two-component model of particle production [34]. In any given centrality, we observed a wide range of nucleon densities in L+R binning from Fig. 2a, b. Therefore, a larger $dN_{ch}/d\eta$ is obtained for events in L+R bins belonging to more central collisions even for a fixed $N_{part}$. In other words, for a given $N_{part}$, it is possible to choose events with larger event activity using L+R binning. It is important to note that $dN_{ch}/d\eta$ is more strongly correlated with $N_{part}$ than with $N_{coll}$ in a collision event [35].[1]

Figure 3a, b shows $\varepsilon_2$ and $v_2$ calculated using the event-plane method as a function of $dN_{ch}/d\eta$. $\varepsilon_2$ is geometry-

---
[1] Further details on the nature of $dN_{ch}/d\eta/N_{part}$ and $dN_{ch}/d\eta/N_{coll}$ as a function of $N_{part}$ for different centralities are shown in Appendix A. The plots show that the difference in centrality and L+R binning for $dN_{ch}/d\eta$ mostly arises from $N_{part}$.

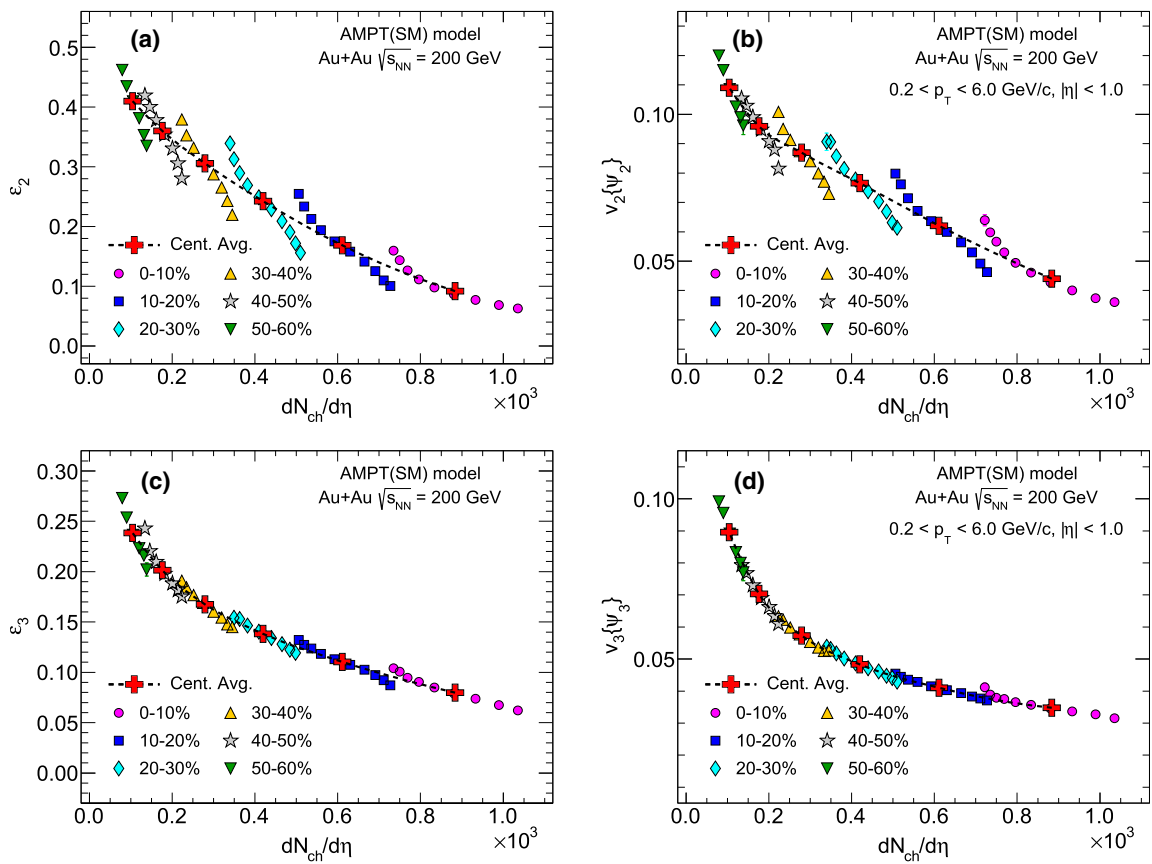

**Fig. 3** Plots for minimum-bias Au+Au collisions at $\sqrt{s_{NN}} = 200$ GeV. Top row: (left) $\varepsilon_2$ vs. $dN_{ch}/d\eta$; (right) $v_2$ vs. $dN_{ch}/d\eta$. Bottom row: (left) $\varepsilon_3$ vs. $dN_{ch}/d\eta$; (right) $v_3$ vs. $dN_{ch}/d\eta$. Centrality average values are shown by the "+" symbol in red, whereas the values for L+R binning in each centrality are shown in solid circles of different colors

driven, and $v_2$ is shown to be proportional to $\varepsilon_2$ following $v_2 = \kappa\varepsilon_2$, where $\kappa$ is the proportionality constant dependent in particular on the equation of state and medium viscosity [36]. Because $v_2$ arises as a hydrodynamic response to $\varepsilon_2$, the overall trend as a function of $dN_{ch}/d\eta$ is expected to be similar. As $\varepsilon_2$ increases from central to peripheral collisions, the magnitude of $v_2$ also increases. $\varepsilon_2$ in L+R binning shows a strong deviation and decreases more sharply than the centrality averaged trend. From Fig. 2a, b, it can be observed that for a given $N_{part}$, a more central event is more dense in L+R binning. This would lead to a more compact collision region, which results in smaller $\varepsilon_2$, as clearly observed in Fig. 3a. Similar behavior is observed for $v_2$ as shown in Fig. 3b, as expected from the linear dependence of $v_2$ on $\varepsilon_2$. In addition, L+R binning can be used to choose events with similar $dN_{ch}/d\eta$ but different $v_2$. This could be particularly helpful in disentangling the effect of initial geometry and initial-state fluctuations on the observed final-state $v_2$. For instance, using L+R binning within a given centrality, we could choose events with smaller area but larger eccentricity (or vice versa) for a given $N_{part}$ and then study how final-state $v_2$ depends on these initial-state parameters. Figure 3b also shows that L+R binning can be used to select events with the same $dN_{ch}/d\eta$ but a higher $v_2$ for more central events. This, in turn, implies that certain anomalous events with higher eccentricity can arise from a more central collision. Having the same $v_2$ across events with very different $N_{ch}$ also provides an important means to separate the contribution from initial-state geometry and fluctuations towards final-state $v_2$. L+R binning could specifically help in studying the contribution from geometric effects such as deformation in nuclear shape, whose dependence differs from factors such as system size, which contributes to fluctuations in final-state $v_2$.

On the other hand, $\varepsilon_3$ is known to be mostly fluctuation-driven [37]. The $\varepsilon_3$ decreases from peripheral to central events owing to decreasing fluctuations [15] as observed in Fig. 3c. The points corresponding to L+R binning for $\varepsilon_3$ versus $dN_{ch}/d\eta$ are closer to the centrality averaged values than those for $\varepsilon_2$. The final-state $v_3$ also displays similar behavior as $\varepsilon_3$, as observed in Fig. 3d. This could also suggest that the initial-state fluctuations contributing to final-state $v_3$ are mostly driven by fluctuations in $b$.

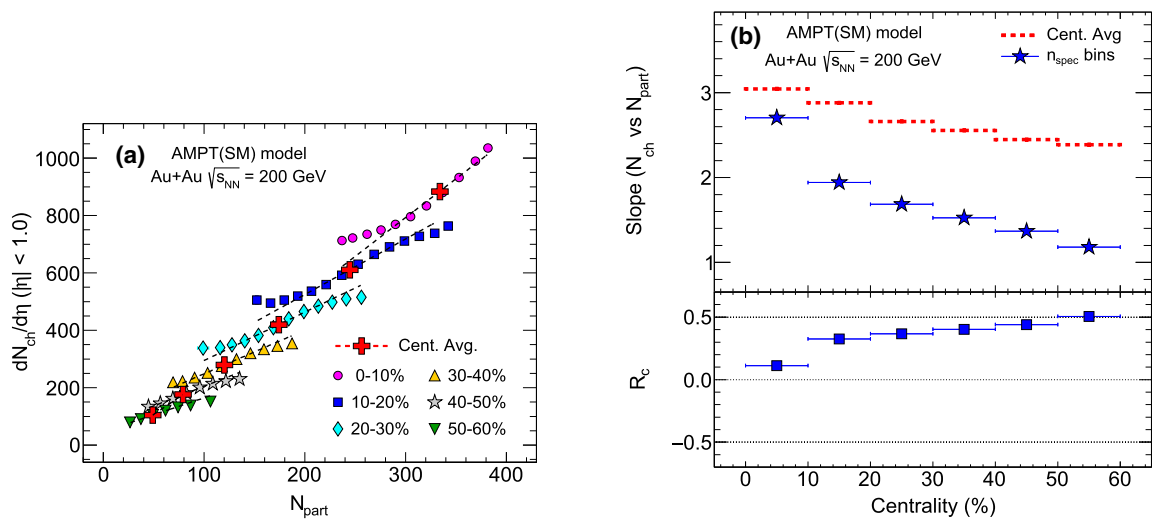

**Fig. 4** Plots showing the method used to quantify the deviation of $n_{\text{spec}}$ binned values from the centrality trend in Au+Au collisions at $\sqrt{s_{NN}}$ 200 GeV. (Left) $dN_{\text{ch}}/d\eta$ vs. $N_{\text{part}}$ showing the linear fits for centrality averaged values and L+R bins for each centrality in dotted lines. (Right) The extracted slope shown as a function of centrality, with the red dotted line showing the slope of the centrality averaged values and solid stars showing the average slope of the values in the $n_{\text{spec}}$ bin for each centrality; the bottom panel shows the relative difference between slopes ($R_c$)

The novelty of this work lies in the quantification of the difference in correlation between observables in centrality and L+R binning to improve access to initial-state configurations. We assume a linear dependence in the binning and also use a linear fit in extracting the slopes. The fitting is performed separately for centrality averaged values and L+R binned values in each centrality. For centrality averaged values, fitting is performed using one point on either side for a given centrality class. The range of fitting in L+R binning is varied over two, three, and four L+R bins on either side of the mean value for each centrality. The convergence of the slopes thus obtained for L+R binning is chosen as the nominal value to infer properties of rare events. We also check the value for the slope parameter, taking the first coefficient from a second-order polynomial fit. The slope thus obtained is observed to be consistent with that obtained from linear fitting. The centrality averaged slope reflects the correlation over a range of centralities, whereas the slope for L+R bins reflects the correlation within a particular centrality. Thus, a relative difference between these two slopes for any given centrality shows the residual correlation in addition to the centrality averaged trend and encodes additional details about the events that are otherwise hidden in the standard centrality binning method. We define a quantity $R_c$ which is the relative ratio between the centrality averaged slope ($S_c$) and the slope for the L+R binning ($S_{ns}$) for each centrality, $R_c = (S_{ns}/S_c) - 1$. If the final-state observable is purely fluctuation-driven, such as impact parameter fluctuations or uncertainties in nucleon positions, the $R_c$ is expected to have a value of zero. The deviation from baseline of zero denotes the extent to which a certain observable is dominated by dynamic effects other than impact parameter fluctuations or the uncertainties in nucleon positions.

The procedure to extract the slopes using linear fitting is shown in Fig. 4a for $dN_{\text{ch}}/d\eta$ versus $N_{\text{part}}$. Figure 4b shows the extracted slopes and $R_c$ as a function of centrality. The extracted slopes for centrality binning are shown in red dotted lines, and those for L+R binning are shown in blue stars. From central to peripheral events, the slope decreases much more rapidly in L+R binning than the slopes from centrality binning. This difference in slopes is directly reflected in the magnitude of $R_c$, as shown in the bottom panel of Fig. 4b. The slopes extracted from the two methods are observed to be closest for the most central events (0–10%). This is expected because of very similar magnitudes of $\rho(n_{\text{spec}}, N_{\text{part}})$ and $\rho(n_{\text{spec}}, N_{\text{ch}})$, as shown in Fig. 1c. Therefore, one can obtain very little additional information on the correlation between $dN_{\text{ch}}/d\eta$ and $N_{\text{part}}$ in L+R binning in comparison with centrality binning. From Fig. 4a, one can observe an almost linear dependence of $dN_{\text{ch}}/d\eta$ with $N_{\text{part}}$ in centrality binning using the AMPT model for Au+Au collisions at $\sqrt{s_{NN}} = 200$ GeV. $R_c$ is observed to increase from central to peripheral collisions. A deviation in $R_c$ from baseline of zero can be interpreted as a deviation from linearity of $dN_{\text{ch}}/d\eta$ in the final state to $N_{\text{part}}$ from the initial state. Therefore, towards more peripheral centralities, the particle production in L+R binning can capture contributions from factors other than those contributing to particle production in just centrality binning. As we go from central to peripheral collisions, the particle production mechanism becomes more complicated

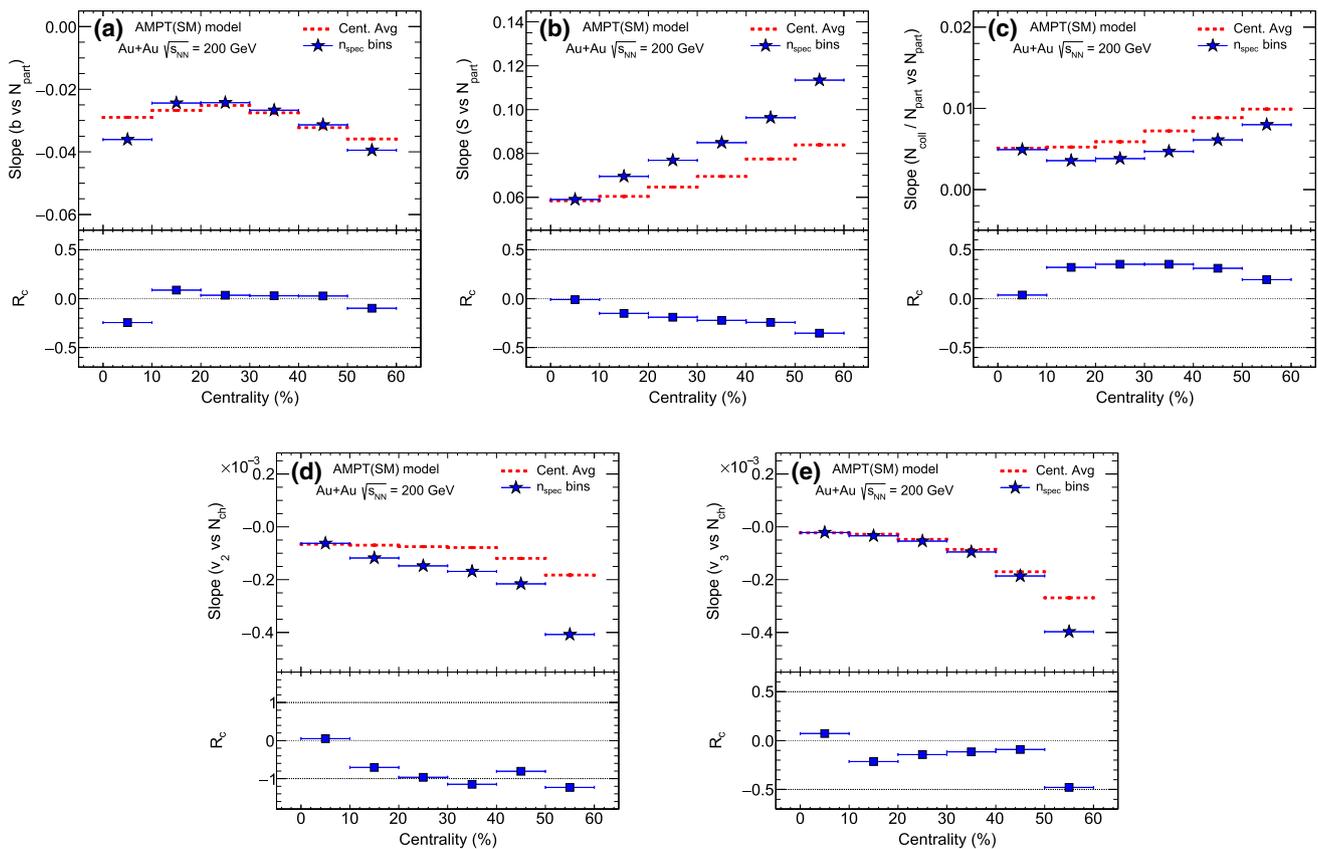

**Fig. 5** Slope for centrality averaged values (red dashed line) and for L+R bins (solid stars) extracted using local linear fits. Top row: (left) slopes for $b$ vs. $N_{part}$; (middle) slopes for $S$ vs. $N_{part}$; (right) slopes for $N_{coll}/N_{part}$ vs. $N_{part}$. Bottom row: (left) slopes for $v_2$ vs. $N_{ch}$; (right) slopes for $v_3$ vs. $N_{ch}$. The bottom panel in all figures shows the relative ratio between slopes; $R_c$ as a function of centrality

than just a linear response to initial $N_{part}$. The multiple binning procedure can help in deciphering more details about the initial-state configuration and energy deposition mechanism to understand the evolution of particle production.

Figure 5a shows slopes for $b$ versus $N_{part}$ in centrality and L+R binning. The slopes for centrality averaged values and L+R binned values are very close for all centralities. This observation can also be seen in Fig. 1c, where for a given $N_{part}$, L+R binning and centrality binning correspond to the same $b$. In other words, because $b$ has one-to-one correspondence with $n_{spec}$, the event selection between centrality and L+R binning is not biased.

Figure 5b shows slopes for $S$ versus $N_{part}$ in centrality and L+R binning. The slopes are positive because increasing $N_{part}$ leads to a larger $S$. The $R_c$ for $S$ versus $N_{part}$ is observed to decrease from central to peripheral collisions. A steady decrease in $R_c$ implies a consistently increasing slope for L+R binning compared with centrality binning. This increasing difference in slopes implies better access to events with larger $S$ with similar $N_{part}$ for more peripheral events. Therefore, using L+R binning, it is possible to access events with larger $S$ or smaller density at the same $N_{part}$ for 10–60% centrality. This could be leveraged to select events with very different initial-state densities at the same $N_{part}$, which could shed light on particle production mechanisms and the contribution of rare initial states to final-state observables.

Figure 5c shows slopes for $N_{coll}/N_{part}$ versus $N_{part}$ in centrality and L+R binning. The slopes are positive, implying an increase in the rate of binary collisions with increasing $N_{part}$. The slope in L+R binning is consistently smaller than the centrality averaged slope, and $R_c$ shows a maximum about mid-centrality. This shows that the rate of binary collision within each centrality has contributions other than just the $N_{part}$. One obvious source could be the $S$, which itself takes a range of values within each centrality and could lead to a larger $N_{coll}/N_{part}$ for smaller $S$ and vice versa (as also shown previously in Fig. 2a). The large magnitude of $R_c$ also implies that L+R binning gives improved access to the events in which the rate of interaction is lower for more peripheral events for a fixed $N_{part}$. In other words, it allows us to study events belonging to the same centrality but with different rates of interactions which could result in different $dN_{ch}/d\eta$ in the final state (also shown in Fig. 2c). Thus, L+R binning can be used to shed light on how the initial rate of interaction ($N_{\text{coll}}/N_{\text{part}}$) in different centralities impacts the final-state particle production mechanism.

Figure 5d, e shows slopes for $v_2$ and $v_3$ versus $N_{\text{part}}$ and their respective $R_c$. $v_2$, being dominated by effects from the shape of initial geometry, shows a larger $R_c$ going from central to peripheral events. If the final-state $v_2$ were driven purely by initial-state fluctuations in participant positions, the slope of $v_2$ versus $N_{\text{part}}$ would have been expected to be the same in centrality and L+R binning. This in turn would have led to an $R_c$ value of zero. The deviation from baseline of zero denotes the extent to which $v_2$ is dominated by initial-state geometry in addition to the initial-state fluctuations. Another interesting point to note is that in most central collisions, as the origin of $v_2$ is dominated by initial-state fluctuations [38], the $R_c$ parameter is close to zero for 0–10% centrality. For $v_3$, the slopes from centrality and L+R binning are very close over a wide range in centrality (0–50%). This reinforces the argument that $v_3$ arises mostly from initial-state fluctuations, and the additional information that could be obtained from L+R binning in this case is quite limited.

## 4 Conclusion

We showed that $n_{\text{spec}}$ could be used as an additional classifier of event characteristics on top of $N_{\text{ch}}$ owing to the very weak correlation between $n_{\text{spec}}$ and $N_{\text{ch}}$. A new method was developed to quantify the difference in the slope of correlations between observables in centrality and L+R binning using $R_c$. A large magnitude of $R_c$, up to 50%, was observed for the correlation between $S$ and $N_{\text{part}}$. This enables us to choose a very different initial overlap area for the same $N_{\text{part}}$. Such a selection over a wide range of overlap area at fixed $N_{\text{part}}$ via L+R binning also leads to widely varying initial-state densities, by about 40%. The resulting variation in initial-state densities and interaction rates also leads to a wide range of $dN_{\text{ch}}/d\eta$ for a given $N_{\text{part}}$. The increasing magnitude of $R_c$ for $dN_{\text{ch}}/d\eta$ versus $N_{\text{part}}$ towards peripheral centrality might point to the nonlinear contributions in initial-state configurations or in the energy deposition mechanism in addition to the linear contribution from $N_{\text{part}}$ in a simple two-component model. The behavior of observables which are known to have dynamic contributions from both initial state and impact parameter fluctuations is very different between centrality averaged and L+R binned values. As a result, we observe a large magnitude of $R_c$ for the correlation between $v_2$ and $N_{ch}$, whereas the $R_c$ for $v_3$ versus $N_{ch}$ is very small, ∼10% on average. Therefore, a separation of the contribution of initial-state geometry from that arising purely from fluctuations could be obtained for $v_2$ using $R_c$.

L+R binning provides a unique means to differentiate between events with widely different initial-state density and rate of binary collisions even at a fixed $N_{\text{part}}$. The effect of such rare and widely varying initial-state configurations within each centrality class could not be accessed by standard centrality binning. In addition, because of the reduced sensitivity of $R_c$ towards initial-state fluctuations, it could also potentially aid in separating the contributions of geometry and initial-state fluctuations towards final-state $v_2$. Such discriminating power of the L+R binning procedure could be leveraged in the future to select events with rare properties to disentangle the impact of several initial-state parameters on final-state observables. In particular, this novel binning procedure could be applied to probe the properties of initial states in heavy-ion collision systems. Data from ZDC detector systems installed at RHIC and the Large Hadron Collider (LHC) could be specifically used to obtain event-by-event spectator neutrons for a more realistic test of this method. Further improvement in the ZDC design in future experiments could help increase the single-neutron peak resolution. Such improvements would also allow us to extend the centrality range in which the multiple binning procedure could be implemented. It would be of great interest to study the implication of L+R binning for smaller or deformed systems which are known to have additional contributions to final-state fluctuation measurement when compared with larger, spherical systems.

**Acknowledgements** SB is supported by U.S. Department of Energy grant number DE-FG02-87ER40331.

**Data Availability Statement** This manuscript has no associated data or the data will not be deposited. [Authors' comment:This work is based on model simulation. The data for generating figures in this work can be provided upon request.]

## Appendix A

Additional plots to study the origin of $dN_{\text{ch}}/d\eta$ from $N_{\text{coll}}$ and $N_{\text{part}}$ are given below (Fig. 6).

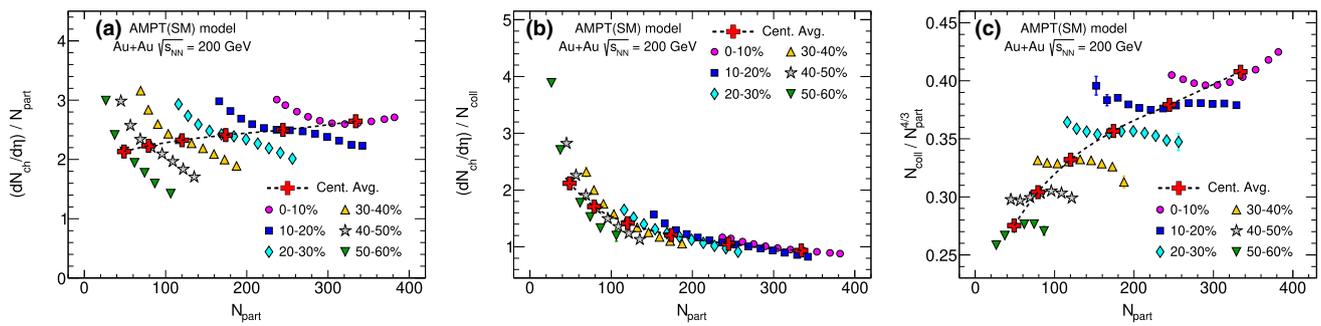

**Fig. 6** Plots for minimum-bias Au+Au collisions at $\sqrt{s_{NN}} = 200$ GeV. (Left) $dN_{ch}/dg\eta/N_{part}$ vs. $N_{part}$; (center) $dN_{ch}/d\eta/N_{coll}$ vs. $N_{part}$; (right) Geometric scaling $N_{coll}/(N_{part})^{4/3}$ vs. $N_{part}$ [12]. Centrality average values are shown by the "+" marker in red, whereas the values for L+R binning in each centrality are shown in solid circles of different colors

### References

1. D. Teaney, E. Shuryak, Phys. Rev. Lett. **86**, 4783–4786 (2001)
2. J. Adams et al., STAR Collaboration, Nucl. Phys. A **757**, 102 (2005)
3. J. Bernhard et al., Phys. Rev. C **94**, 024907 (2016)
4. D. Everett et al., JETSCAPE Collaboration, Phys. Rev. C **103**, 054904 (2021)
5. G. Nijs et al., Phys. Rev. Lett. **126**, 202301 (2021)
6. H.J. Drescher et al., Phys. Rev. C **76**, 024905 (2007)
7. P. Romatschke, U. Romatschke, Phys. Rev. Lett. **99**, 172301 (2007)
8. M. Luzum, P. Romatschke, Phys. Rev. C **78**, 034915 (2008) [Erratum: Phys. Rev. C **79**, 039903 (2009)]
9. V. Roy et al., Phys. Rev. C **86**, 014902 (2012)
10. H. Song et al., Phys. Rev. Lett. **106**, 192301 (2011) [Erratum: Phys. Rev. Lett. **109**, 139904 (2012)]
11. V. Roy et al., J. Phys. G **40**, 065103 (2013)
12. M.L. Miller et al., Annu. Rev. Nucl. Part. Sci. **57**, 205–243 (2007)
13. C. Loizides, Phys. Rev. C **94**, 024914 (2016)
14. A.M. Poskanzer, S.A. Voloshin, Phys. Rev. C **58**, 1671 (1998)
15. B. Alver, G. Roland, Phys. Rev. C **81**, 054905 (2010) [Erratum: Phys. Rev. C **82**, 039903 (2010)]
16. C. Gale et al., Int. J. Mod. Phys. A **28**, 1340011 (2013)
17. U. Heinz, R. Snellings, Annu. Rev. Nucl. Part. Sci. **63**, 123 (2013)
18. F.G. Gardim et al., Phys. Rev. C **85**, 024908 (2012)
19. J. Xu, C.M. Ko, Phys. Rev. C **84**, 044907 (2011)
20. J. Xu, C.M. Ko, Phys. Rev. C **84**, 014903 (2011)
21. V. Bairathi et al., Phys. Rev. C **91**, 054903 (2015)
22. C. De Martin (ALICE Collaboration), Quark Matter (2022). https://indico.cern.ch/event/895086/contributions/4736386/. Accessed 12 Sep 2022
23. S. Sombun et al., J. Phys. G **45**, 025101 (2018)
24. V. Bairathi et al., Phys. Lett. B **754**, 144–150 (2016)
25. Xu. Yi-Fei et al., Nucl. Sci. Tech. **27**, 126 (2016)
26. E.G. Judd et al., Nucl. Instrum. Meth. A **902**, 228 (2018)
27. Z.W. Lin et al., Phys. Rev. C **72**, 064901 (2005)
28. X.-N. Wang, M. Gyulassy, Phys. Rev. D **44**, 3501–3516 (1991)
29. B. Zhang, Comput. Phys. Commun. **109**, 193 (1998)
30. B. Andersson et al., Z. Phys. C **20**, 317 (1983)
31. J. Adam et al., ALICE Collaboration, JHEP **2015**, 97 (2015)
32. ATLAS Collaboration, arXiv:2205.00039 [nucl-ex]
33. C. Loizides et al., SoftwareX **1–2**, 13–18 (2015)
34. D. Kharzeev, M. Nardi, Phys. Lett. B **507**, 121 (2001)
35. J.S. Moreland et al., Phys. Rev. C **92**, 011901 (2015)
36. R.S. Bhalerao et al., Phys. Rev. C **84**, 034910 (2011)
37. GKh. Eyyubova et al., J. Phys. G **48**, 095101 (2021)
38. H. Petersen et al., Phys. Rev. C **82**, 041901 (2010)